\documentclass[preprint,12pt]{elsarticle}

\usepackage{amssymb}
\usepackage{amsthm}
\usepackage{xcolor}

\usepackage[utf8]{inputenc}
\usepackage[T1]{fontenc}
\usepackage{amsfonts}
\usepackage{amsmath}
\usepackage{subcaption}
\usepackage{array}
\usepackage{hyperref}
\usepackage{booktabs}

\usepackage{tikz}
\usetikzlibrary{shapes.geometric, arrows}

\tikzstyle{startstop} = [rectangle, rounded corners, 
minimum width=3cm, 
minimum height=1cm,
text centered, 
draw=black, 
fill=gray!30]

\tikzstyle{io} = [trapezium, 
trapezium stretches=true, 
trapezium left angle=70, 
trapezium right angle=110, 
minimum width=3cm, 
minimum height=1cm, text centered, 
draw=black, fill=gray!30]

\tikzstyle{process} = [rectangle, 
minimum width=3cm, 
minimum height=1cm, 
text centered, 
text width=3cm, 
draw=black, 
fill=gray!30]

\tikzstyle{decision} = [diamond, 
minimum width=3cm, 
minimum height=1cm, 
text centered, 
draw=black, 
fill=gray!30]
\tikzstyle{arrow} = [thick,->,>=stealth]

\journal{Computational Economics}

\begin{document}

\begin{frontmatter}



\title{Technical Analysis Meets Machine Learning: Bitcoin Evidence}

\author[1]{José Ángel Islas Anguiano\corref{equal}\fnref{orcidA}}
\ead{islas.jose@uas.edu.mx}

\author[2]{Andrés García-Medina\corref{equal}\fnref{orcidB}}
\ead{andgarm.n@gmail.com}

\address[1]{Facultad de Ciencias Físico-Matemáticas, Universidad Autónoma de Sinaloa, 
Av. de las Américas y Blvd. Universitarios S/N, Culiacán, Sinaloa CP 80010, Mexico}

\address[2]{Independent Researcher, Ensenada 22840, Mexico}

\cortext[equal]{These authors contributed equally to this work.}
\fntext[orcidA]{ORCID: 0000-0001-5349-775X}
\fntext[orcidB]{ORCID: 0000-0002-2198-880X}

\begin{abstract}
In this note, we compare Bitcoin trading performance using two machine learning models—Light Gradient Boosting Machine (LightGBM) and Long Short-Term Memory (LSTM)—and two technical analysis-based strategies: Exponential Moving Average (EMA) crossover and a combination of Moving Average Convergence/Divergence with the Average Directional Index (MACD+ADX). The objective is to evaluate how trading signals can be used to maximize profits in the Bitcoin market. This comparison was motivated by the U.S. Securities and Exchange Commission’s (SEC) approval of the first spot Bitcoin exchange-traded funds (ETFs) on 2024-01-10.
Our results show that the LSTM model achieved a cumulative return of approximately 65.23\% in under a year, significantly outperforming LightGBM, the EMA and MACD+ADX strategies, as well as the baseline buy-and-hold. This study highlights the potential for deeper integration of machine learning and technical analysis in the rapidly evolving cryptocurrency landscape.

\end{abstract}




\noindent\textbf{Preprint notice.} This manuscript is a \emph{preprint} submitted for consideration in a Springer journal. 
It has not been peer reviewed. The content may change before final publication. 
Once the article is published in its final form, this version will be updated with a reference to the published article, including its DOI.
\medskip

\end{frontmatter}


\section{Introduction}


Machine learning is widely used across science and engineering to extract patterns from data and identify similar structures in other datasets. In finance, various techniques—such as neural networks and regression—have been applied to support decision-making, though their suitability is sometimes debated. In particular, machine learning (ML) can be applied to financial trading data to train regression models on daily closing prices, aiming to detect conditions for buying or selling positions. On the other hand, technical analysis (TA), often considered a heuristic tool, is widely used to identify buy or sell signals in financial assets such as commodities or stocks. These tools are based on empirical observations rather than formal mathematical models.


In this note, we employ two technical analysis (TA) strategies: Exponential Moving Average (EMA) crossovers and a combination of Moving Average Convergence/Divergence with the Average Directional Index (MACD+ADX). The machine learning (ML) methods applied are Light Gradient Boosting Machine~(LightGBM)~\cite{ke2017lightgbm} and the Recurrent Neural Network model (RNN) named Long Short-Term Memory~(LSTM)~\cite{hochreiter1997long}. The primary objective of this work is to compare the trading signals generated by TA and ML approaches in terms of cumulative returns on Bitcoin trading. To ensure a fair comparison under realistic market conditions, both approaches are integrated into a statistical learning framework in which model parameters are optimized in-sample and evaluated out-of-sample.


The purpose is to use the trading strategies to maximize profits in the Bitcoin digital commodity. Both trading approaches use the same dataset for training and a separate dataset for testing. The training dataset consists of price data from 2021-01-10, to 2024-01-09, while the testing dataset covers the period beginning with the approval of the first spot Bitcoin exchange-traded funds (ETFs) by the U.S. Securities and Exchange Commission (SEC)\footnote{\url{https://www.cftc.gov/PressRoom/PressReleases/7231-15}} on 2024-01-10, and ending on 2024-12-31. It is important to note that, in order to establish a fair comparison between technical analysis and machine learning approaches, the former does not require a window period to begin generating predictions, whereas the latter necessitates an initial 95-day window period to produce forecasts on the testing data. Consequently, the returns are calculated for the period from 2024-04-14, to 2024-12-31.


One of the main objectives of this work is to address a gap in the literature by examining the effects of shifting market dynamics triggered by the release of the Bitcoin ETF. On the one hand, we evaluate the performance of trading strategies trained prior to this structural change and tested afterward, during the artificial turbulence generated by this economic event.
Despite the shift in dynamics, the results show surprisingly high cumulative returns across all trading strategies. In particular, the LSTM-based trading strategy outperforms the classical buy-and-hold approach typically followed by fundamental investors~\cite{graham2003intelligent}.

In Section 2, we present a brief literature review. Section 3 describes the LightGBM and LSTM models, as well as the technical analysis strategies implemented in this paper. Section 4 presents the main results of this work, while Section 5 provides a discussion of these results. Finally, Section 6 offers a general conclusion and mentions possible venues for future work.

\section{Literature Review}

Bitcoin has traditionally been considered a speculative asset~\cite{baur2018bitcoin}. However, recent studies have shown that its behavior more closely resembles that of a commodity due to its fixed supply. In~\cite{gronwald2019bitcoin}, Bitcoin was compared to oil and gold through volatility models, arguing that its dynamics are similar, although showing that Bitcoin exhibits more extreme dynamics due to demand shocks. Nevertheless, there is no clear and definitive consensus. The authors of~\cite{white2020bitcoin} question and deeply debate whether Bitcoin can be considered an asset, a currency, a commodity, a technological product, or something else. New arguments have emerged, suggesting that the cost of mining Bitcoin should be considered in determining whether it qualifies as a commodity. In~\cite{baldan2020bitcoin}, this aspect is addressed, finding that the effect is not sufficient to fully explain Bitcoin’s dynamics, and suggesting that new drivers should be integrated. In~\cite{rotta2022bitcoin}, the authors go beyond these classifications and define Bitcoin under a new conceptual category: a \emph{digital commodity}, implying that it is a commodity generated through human labor, but constrained to the world of bits and bytes, without any real added value.
Nonetheless, beyond this interesting discussion, in this work we focus on Bitcoin in terms of profitability within trading strategies.

Traditionally, technical analysis serves as an auxiliary tool for determining optimal moments to buy or sell financial stocks. This trading strategy involves evaluating a set of assumptions about market trends to guide investment decisions.
Efforts have been made to integrate technical analysis with other areas of investment management, such as portfolio theory. For instance, Santos et al.~\cite{santos2022markowitz} reconcile Markowitz's theory with strategies based on technical indicators. Their approach extends technical analysis beyond timing decisions, aiming to optimize both the quantity of shares or capital allocated to each transaction and the timing of buy or sell actions.
Nti et al. \cite{nti2020systematic} present an excellent review of one hundred and twenty-two (122) studies conducted between 2007 and 2018 related to machine learning strategies for financial market prediction. The review specifically focuses on approaches based on technical analysis, fundamental analysis, and the combination of both schools of thought.
A study covering 60 years of historical data from the FT30 index has shown that simple technical analysis strategies, such as the Moving Average Convergence–Divergence (MACD) and the Relative Strength Index (RSI), can outperform the classic buy-and-hold paradigm typically followed by fundamental analysis practitioners\cite{chong2008technical}.
These strategies are also examined across five indices from OECD countries, likewise showing that they can outperform the basic buy-and-hold strategy under certain specific parameter settings of the indicators \cite{chong2014revisiting}.


Li et al. \cite{li2020stock} provide a review of various attempts to integrate technical analysis into financial time-series forecasting. In the reviewed works, technical analysis signals are used as inputs (features) to predictive models. The study by Singh et al. \cite{singh2022application} also falls into this category, where technical indicators are used as input to an LSTM model to forecast the S\&P 500 index.
A similar attempt to enhance technical analysis through machine learning has been applied to the sectors comprising the S\&P 500. However, two critical limitations of that work are worth noting: it does not compare performance against the traditional buy-and-hold strategy, nor does it explicitly define how the algorithm learns from the technical analysis outcomes.
Moreover, the study by Macedo et al. \cite{macedo2020comparative} proposes a hybrid model that applies genetic algorithms to identify the most suitable technical indicator to follow. However, this approach, which is more closely aligned with operations research, is not adopted in the present work. In contrast, our objective is to compare the strengths of both predictive modeling and classical technical analysis.

Recently, there has been growing interest in applying trading strategies to the cryptocurrency market due to its arbitrage opportunities, which have been shown to be greater than those in traditional markets. Fang et al. \cite{fang2022cryptocurrency} discuss the approaches adopted in recent years by reviewing a set of 146 papers, which implement both econometric and machine learning techniques in the development of trading strategies.
In this emerging line of research, there are also attempts to combine machine learning methods with classical technical analysis. In \cite{goutte2023deep}, technical indicators are used as features within LSTM and GRU neural network architectures, demonstrating superior performance compared to the buy-and-hold baseline approach.
The authors of~\cite{grobys2020technical} have also extended the application of technical trading strategies to the context of cryptocurrencies. In their study, they analyze 10 altcoins, excluding Bitcoin, and test the Variable Moving Average (VMA) oscillator strategy. Their results are promising, paving the way for further extensions in this area.
Strategies more closely related to the fields of optimization and automated trading have also been proposed. In \cite{omran2023optimization}, a new normalized decomposition-based multi-objective particle swarm optimization (N-MOPSO/D) algorithm is implemented, demonstrating strong performance in terms of Return on Investment (ROI), Sortino Ratio (SOR), and the number of trades (TR). However, the comparative analysis is limited, as it does not include the buy-and-hold baseline based on cumulative returns.


On the other hand, in financial time series analysis, traditional statistical and econometric approaches often face challenges when modeling non-stationary variables or capturing complex dependencies \cite{dixon2015implementing}. Fortunately, deep learning techniques are well-suited to identifying and managing these intricate patterns \cite{fischer2018deep}.
Garcia-Medina and Aguayo-Moreno \cite{garcia2024lstm} investigate this approach to forecast the volatility of various cryptocurrencies within the framework of a hybrid LSTM-GARCH model. Similarly, Garcia-Medina and Toan \cite{garcia2021drives} examine the main determinants of Bitcoin prices using data mining techniques and a predictive classification model based on LSTM.
On the other hand, LightGBM was the top-performing model in the M5 forecasting competition focused on business time series \cite{makridakis2022m5, makridakis2022m52}, in terms of both accuracy and uncertainty.


In summary, this study aims to evaluate a series of trading strategies in the context of Bitcoin analysis. Our primary contribution lies in the integration of these techniques into a statistical learning framework, where model parameters are optimized in-sample and applied out-of-sample. This approach enables a fair comparison of indicators under realistic market conditions. Specifically, we compare the performance of technical trading strategies with a method to forecast the direction of Bitcoin prices, based on the methodology presented in previous studies by one of the authors~\cite{garcia2021drives, garcia2024lstm}. Beyond the integration of technical and forecasting-based strategies, a key contribution of this study is the evaluation of model performance under different market conditions for the underlying asset, induced by the creation of the Bitcoin ETF.

\section{Materials and Methods}
\subsection{Data}
Raw price data were downloaded from the free Binance source\footnote{\url{https://www.binance.com}}. As mentioned in the Introduction, the study period extends from January 19, 2021, to December 31, 2024. The subperiod from April 14, 2024, to December 31, 2024, is employed to compare returns, accounting for the 95-day window required by the machine learning algorithms.

\subsection{LSTM}

The LSTM model, introduced by~\cite{hochreiter1997long}, was designed to handle long-term dependencies in sequential data. It uses special components called \textit{gates} to control the flow of information at each time step, allowing the network to retain only the relevant information for future predictions. This structure addresses common issues in training traditional RNNs, particularly with gradient computation. Unlike standard RNNs, each LSTM unit contains several internal layers: a memory cell~($c_t$), an input gate~($i_t$), a forget gate~($g_t$), and an output gate~($o_t$).

\begin{figure}[hbt]
\centering 
\includegraphics[width=0.8\linewidth]{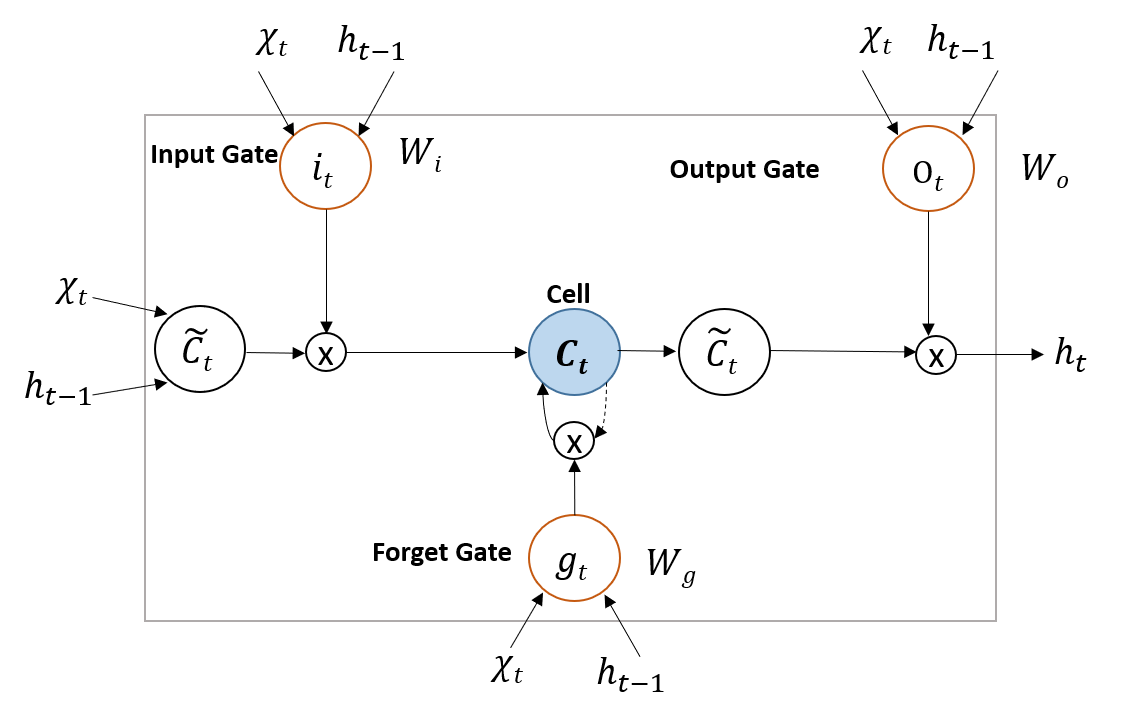} 
\caption{General structure of an LSTM unit\cite{garcia2021drives, garcia2024lstm}. The memory gate is represented as $c_t$, the input gate is denoted by $i_t$,  the forget gate is indicated as $g_t$, and the output gate is as $o_t$.}
\label{fig:02-lstm}
\end{figure}

Figure~\ref{fig:02-lstm} illustrates the general structure and functioning of an LSTM unit.
This diagram illustrates the input $x_t$, the hidden state $h_t$, and $\tilde{c}_t$ at time $t$, which controls the quantity of information retained or discarded in the cell state. The functioning of an LSTM unit can be mathematically represented by the  following equations:

\begin{eqnarray}
   g_t = \sigma(U_g x_t + W_g h_{t-1} + b_f)\\
   i_t = \sigma(U_i x_t + W_i h_{t-1} + b_i)\\
   \tilde{c}_t = tanh(U_x x_t + W_c h_{t-1}+b_c)\\
   c_t = g_t * c_{t-1} + i_t * \tilde{c}_t\\
   o_t = \sigma (U_o x_t + W_o h_{t-1} b_o)\\
   h_t = o_t * tanh(c_t).
\end{eqnarray}

Here, $U$ and $W$ represent weight matrices, $b$  the bias term, and the symbol * denotes element-wise multiplication\cite{kim2018garhlstm,garcia2024lstm}.

\subsection{LightGBM}

The $\mathtt{LightGBM}$ model was proposed in \cite{ke2017lightgbm} to improve the tradeoff of the Gradient Boosting Decision Tree (GBDT)\cite{friedman2001greedy} in the context of accuracy and efficiency.
GBDTs are decision trees that are grown sequentially with the intention of correcting errors or residuals from the predecessor trees. The way to control the fit is through the gradient of a loss function\cite{james2013introduction}.
Essentially, a decision tree is composed of a terminal node, leaf, internal node, and branches. The intention is to recursively determine the binary splitting. This represents a problem in terms of computational cost, since all instances must initially be scanned to estimate the information gained at the split points. The $\mathtt{LightGBM}$ model attempts to solve this problem by integrating Gradient-based One-Side Sampling (GOSS) and Exclusive Feature Bundling (EFB). The first technique involves preserving only the instances with the highest gradient, since these contribute significantly to the information gain. The second improvement integrates a greedy algorithm to reduce the number of features or regressors in the classification problem.

\subsection{Classification Metrics}
In classification tasks, predictions can be compared to the actual class labels. The outcomes of these comparisons are summarized in the confusion matrix, which includes four possible scenarios: true positives (TP), true negatives (TN), false positives (FP), and false negatives (FN).
The following metrics are implemented to quantify the quality of the classification forecast in the price of bitcoin. 

\begin{itemize}
    \item Accuracy = $\frac{TP+TN}{TP+TN+FP+FN}$
    \item Sensitivity, recall or true positive rate~(TPR) $=\frac{TP}{TP+FN}$
    \item Specificity, selectivity or true negative rate~(TNR) $=\frac{TN}{TN+FP}$
    \item Precision or Positive Predictive Value~(PPV) $= \frac{TP}{TP+FP}$
    \item False Omission Rate~(FOR) $= \frac{FN}{FN+TN}$
    \item Balanced Accuracy~(BA) $= \frac{TPR+TNR}{2}$
    \item F1 score $= 2 \frac{PPV\times TPR}{PPV + TPR}$.
\end{itemize}

\subsection{Bitcoin's price direction}

The task of detecting Bitcoin’s price direction was approached using LSTM and $\mathtt{LightGBM}$ models. The first step involved splitting the data into training, validation, and test datasets. The selected training period extends from 2021-01-01, to 2024-01-09, while the test period covers 2024-01-10, to 2024-12-31. Furthermore, a sequential split of 80\% for training and 20\% for validation was applied within the original training period (independent of the test dataset). The original data were then transformed into a supervised learning format, where each input sample consists of 95 consecutive historical days used to predict the following day.
Because we are interested in predicting the direction of Bitcoin (BTC), the time series is not demeaned. Instead, returns are computed and labeled with a value of 1 if positive, and 0 if negative. In this way, the deep learning model is fed a binary time series of zeros and ones, representing upward and downward movements in the raw price of Bitcoin.

\begin{figure}[hbtp]
    \centering
    \includegraphics[width=0.65\linewidth]{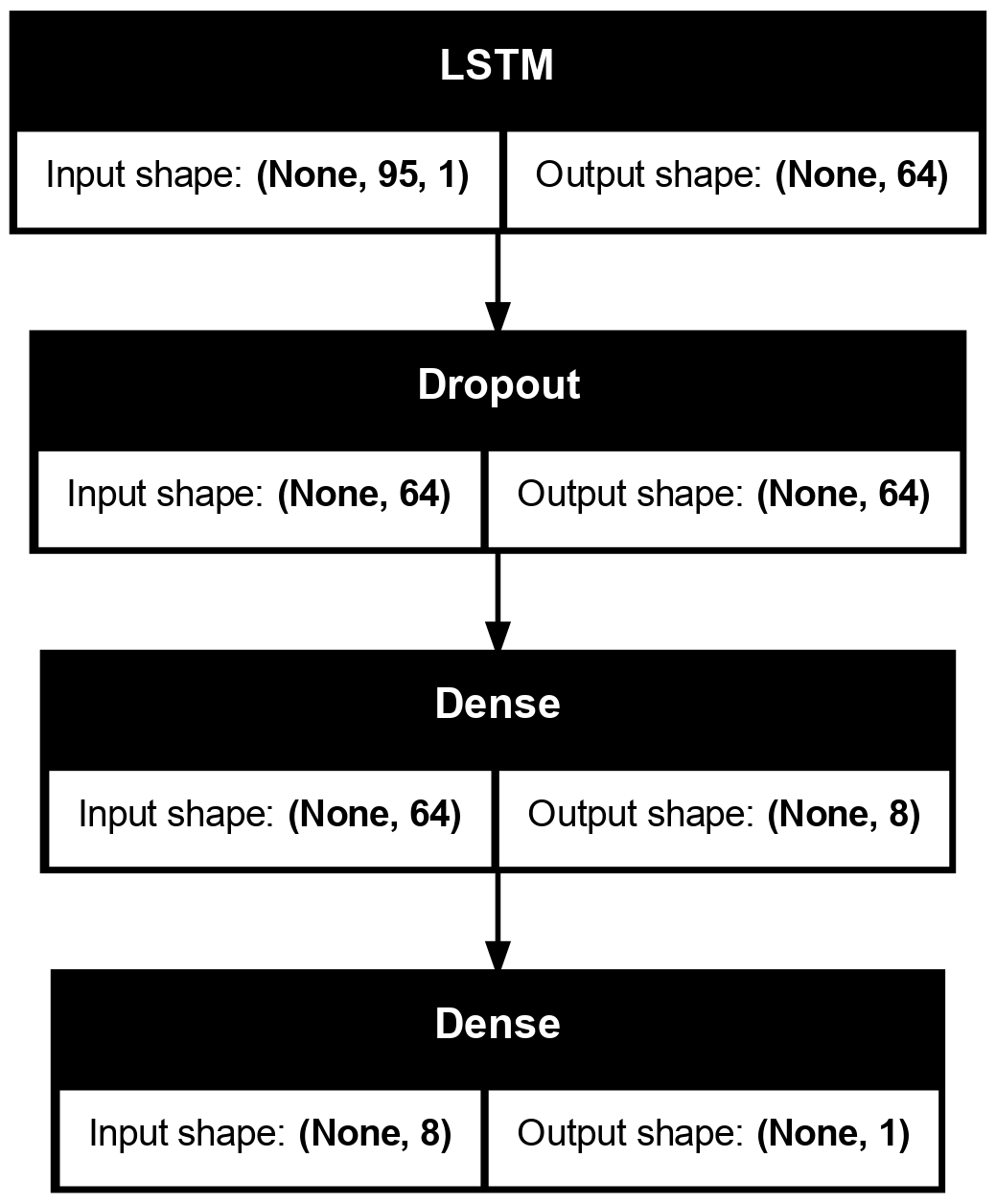}
    \caption{Model architecture of LSTM.}
    \label{modelo_arquitectura}
\end{figure}

In the LSTM model, we use a sigmoid activation function in the output layer, as the task involves binary classification of price direction (upward or downward). Additionally, we implement the variant of the Adam optimizer proposed by~\cite{reddi2019convergence}, which incorporates long-term memory of past gradients to enhance convergence properties.
The number of epochs—i.e., the number of times the learning algorithm passes through the entire training dataset—is determined using an early stopping procedure. The batch size is selected from a predefined grid: {32, 64, 128, 256}. We also explore two learning rates, {0.001, 0.0001}, and apply dropout regularization with rates in {0.3, 0.5, 0.7}. The model was trained by minimizing the binary cross-entropy loss function.
Due to the stochastic nature of deep learning models, we set a random seed and report the average results across ten different random initializations of the parameters to ensure robustness.
The proposed architecture of the neural network is illustrated in Figure~\ref{modelo_arquitectura}, where a dropout layer is included for regularization, along with a final dense layer that produces the binary forecast.

In the $\mathtt{LightGBM}$ model, we generate a grid search to estimate the optimal number of leaves in the set:\{31, 64, 128\}.  Here, we explore three learning rates, {0.1, 0.05, 0.01}. On the other hand, the \emph{bagging} fraction, i.e., the percentage of randomly selected observations to train each tree, was searched over the set \{0.6, 0.8. 1\}. Moreover, two regularization parameters were considered to stabilize the weight of the leaves. We search over \{0, 0.1, 1\} and \{0,0.1,1\} on LASSO (least absolute shrinkage and selection operator) and Ridge regularization~\cite{hastie2009elements}, respectively.
Finally, the model was trained by minimizing the area under the curve (AUC) of the receiver operating characteristic (ROC).
Again, due to the stochastic nature of deep learning models, we set a random seed and report the average results across ten different random initializations of the parameters to ensure robustness.

\subsection{Technical analysis: EMA cross-strategy}

The EMA Cross strategy is a trend-following trading method that uses two exponential moving averages (EMAs) of different lengths to generate buy and sell signals based on momentum shifts. The two EMAs are computed using the closing prices of the asset, in this case, Bitcoin. Typically, a shorter-term EMA (e.g., 12-period) and a longer-term EMA (e.g., 26-period) are used. A buy signal occurs when the shorter-term EMA crosses above the longer-term EMA, indicating a potential upward trend. Conversely, a sell signal is triggered when the shorter-term EMA crosses below the longer-term EMA, suggesting a downward trend.

The strategy is optimized through a grid search over various combinations of short and long EMA window lengths, selecting the pair that yields the highest cumulative return on the training data. As with the LSTM forecasting approach, historical Bitcoin price data from 2021-01-01 to 2024-01-09 is used for training. The strategy is then evaluated on the out-of-sample period from 2024-01-10 to 2024-12-31. To align with the LSTM model’s requirement of a 95-day window before generating forecasts, the trading strategy begins effectively on 2024-04-14.


\subsection{Technical analysis: MACD+ADX}

The MACD + ADX strategy is a hybrid approach that combines two technical indicators to enhance signal reliability. The MACD  detects momentum and potential trend reversals by comparing two EMAs of the asset’s price. A buy signal is generated when the MACD line crosses above the signal line, and a sell signal is generated when it crosses below. The ADX measures the strength of a trend; values above 25 typically indicate a strong trend, while values below suggest a weak or non-trending market. The strategy generates a trade signal only when both indicators align—for example, a MACD buy signal confirmed by a strong ADX reading.

Parameter optimization is performed using a grid search over ranges for the MACD short EMA (e.g., 5–20), MACD long EMA (e.g., 10–30), signal line EMA (e.g., 5–15), and ADX window size (e.g., 10–20). Each parameter combination is evaluated on historical performance, and the one with the highest return is selected as optimal. As with the EMA strategy, the model is trained on Bitcoin data from 2021-01-19 to 2024-01-09 and tested on data from 2024-01-10 to 2024-12-31. Again, to align with the LSTM model’s requirement of a 95-day window before generating forecasts, the trading strategy begins effectively on 2024-04-14.

It is important to mention that these trading strategies do not include transaction costs (brokerage fees, slippage, etc.), which significantly impact real-world performance. Neither uses risk management features like stop-loss orders.

\section{Results}

The comparative cumulative returns of the different strategies are summarized in Table~\ref{tab:cumulative_return}. Among them, the LSTM strategy achieved the highest cumulative return. This strong performance may be attributed to its classification accuracy of 0.5611, as shown in Table~\ref{classification_metrics}, which is notably above the baseline of 0.5. The Buy \& Hold strategy tends to perform well during sustained uptrends, while the MACD+ADX strategy outperforms the EMA-based strategy by incorporating trend strength into its trading signals.

\begin{table}[hbtp]
\centering
\caption{Cumulative return for each trading strategy.}
\label{tab:cumulative_return}
\begin{tabular}{l | c}
\hline
\textbf{Case} & \textbf{Cumulative Return} \\
\hline
LSTM & 65.23\% \\
LightGBM & 53.38\% \\
Buy \& Hold & 42.51\%\\
MACD + ADX & 35.45\% \\
EMA & 26.07\%\\
\hline
\end{tabular}
\end{table}

\begin{table}[hbtp]
\centering
\caption{Classification Metrics}
\label{classification_metrics}
\begin{tabular}{lrr}
\toprule
 & LSTM & LightGBM \\
Metric &  \\
\midrule
Accuracy & 0.5611 & {\bf 0.5840}\\
TPR & 0.4887 & {\bf 0.5639}\\
TNR & {\bf 0.6357} & 0.6047\\
PPV & 0.5804 & {\bf 0.5952}\\
FOR & {\bf 0.4533} &  0.4265\\
BA & 0.5622 & {\bf 0.5843}\\
F1 Score & 0.5306 & {\bf 0.5792}\\
\bottomrule
\end{tabular}
\end{table}

\begin{figure}[hbtp]
\centering
\includegraphics[scale=0.5]{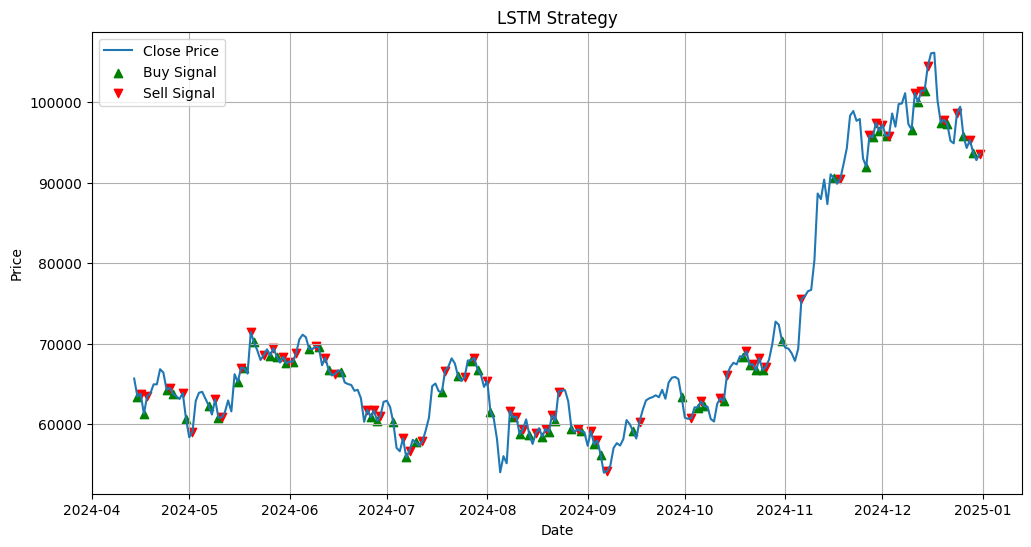}
\caption{LSTM.\label{figLSTM}}
\end{figure}

\begin{figure}[hbtp]
\centering
\includegraphics[scale=0.4]{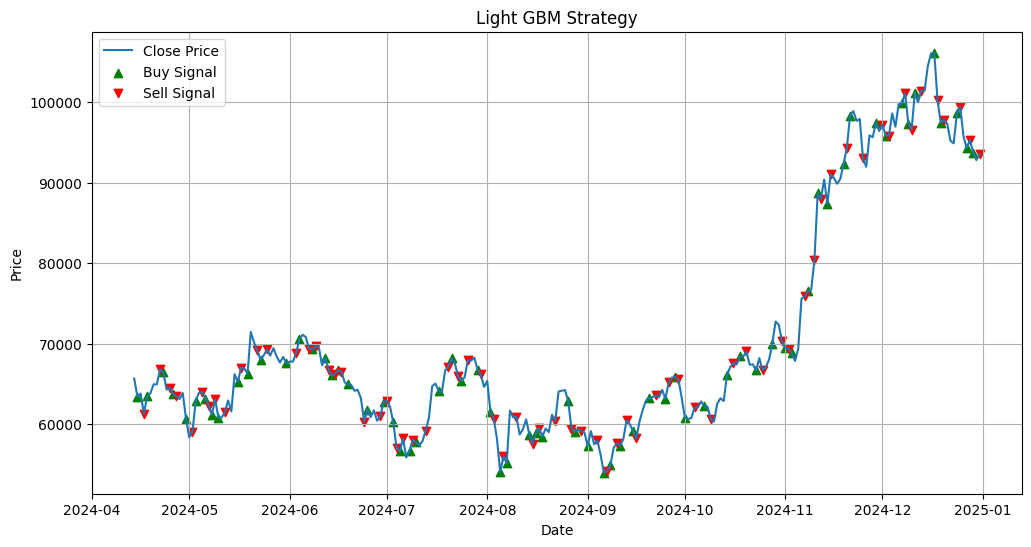}
\caption{Light GBM.\label{figlightGBM}}
\end{figure}

Figure \ref{figLSTM} displays the performance of the optimized LSTM-based trading strategy, while \ref{figlightGBM} te corresponding for the $\mathtt{LightGBM}$ model. This approach generates a high frequency of trades, reflecting the model’s high sensitivity to short-term price fluctuations. The strategy seeks to capitalize on small price movements, often taking profits over relatively short holding periods. Although this can lead to increased trading activity and the potential for quick gains, it can also result in higher exposure to market noise.

\begin{figure}[hbtp]
\centering
\includegraphics[scale=0.4]{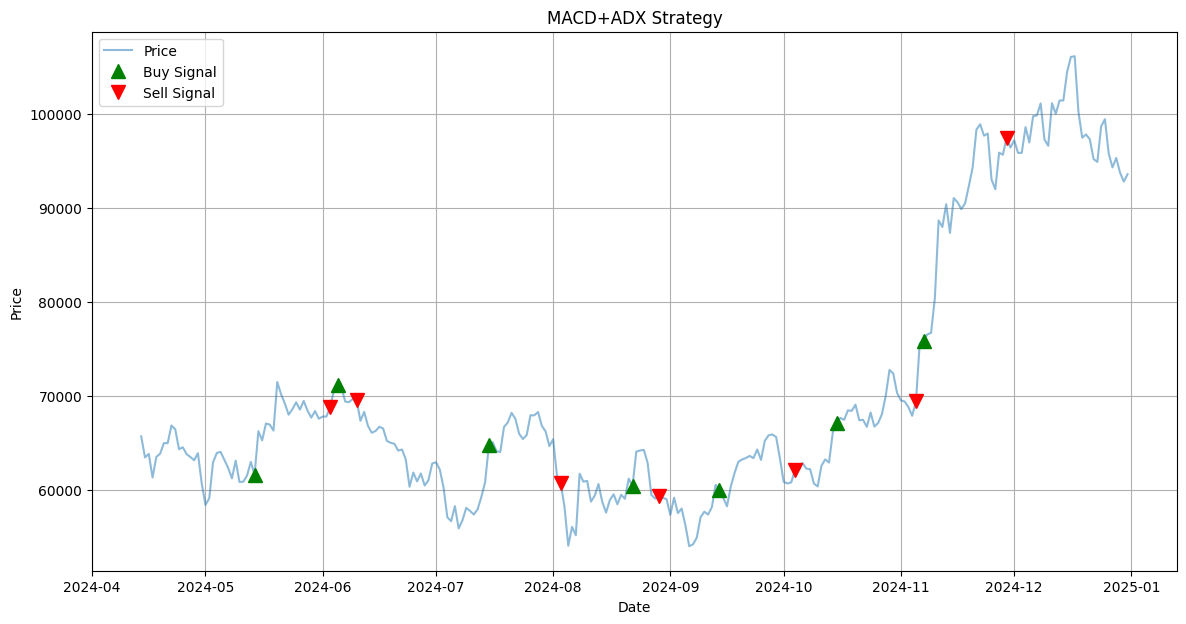}
\caption{MACD+ADX.\label{figMACD}}
\end{figure}

Figure \ref{figMACD} illustrates the execution of trades based on the optimized MACD + ADX strategy. The optimal parameters identified are as follows: MACD Short = 17, Long = 21, Signal = 15, and ADX = 13. The figure shows that the price of Bitcoin gains significant upward momentum from October to December, a trend effectively captured by the strategy due to its incorporation of price strength via the ADX component. The strategy triggers a limited number of trades during this period, focusing on quality over quantity and aligning well with strong directional movements in the market.

\begin{figure}[hbtp]
\centering
\includegraphics[scale=0.4]{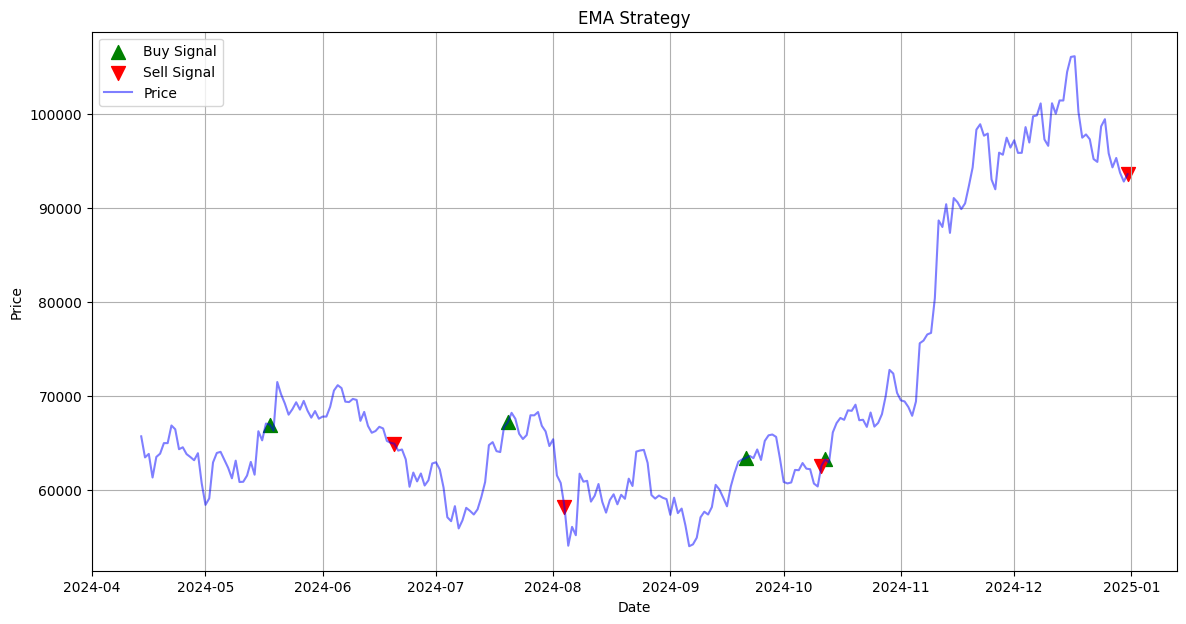}
\caption{EMA.\label{figEMA}}
\end{figure}

Figure \ref{figEMA} presents the execution of the trades using the optimized EMA crossover strategy. The strategy employs a six-day period for the short-term EMA and a 95-day period for the long-term EMA. A buy signal is generated when the short-term EMA crosses above the long-term EMA, indicating a potential upward trend. In contrast, a sell signal, used to close the previous long position, is activated when the short-term EMA crosses below the long-term EMA. Similarly to the MACD + ADX approach, this strategy results in a relatively low number of trades, focusing on capturing significant market movements rather than frequent signals.

\subsection{Further analysis and Trading cost}
Table \ref{table_analysis} summarizes the numbers of trades, the numbers of winning trades, the loss trades, and the profits using a 0.1\% fee for each trade.

\begin{table}[hbtp]
\centering
\caption{Table Analysis.}
\label{table_analysis}
\begin{tabular}{l | c | l| l | c }
\hline
\textbf{Case} & \textbf{\# Trades} &\textbf{Cumulative Return with 0.1\% fee} \\
\hline
LSTM & 120  & 53.23\%\\
LightGBM & 136 & 39.78\% \\
Buy \& Hold &2 & 42.31\% \\
MACD + ADX & 14& 34.45\%\\
EMA & 8 & 25.27\%\\

\hline
\end{tabular}
\end{table}

The formula \cite{Chan2021} for the commission cost $\alpha=0.1\%$ for each buy or sell is  
\[
\text{Total commission cost} = (\text{Number of trades}) (\alpha)
\]
Thus, 
\[
\text{Cumulative Return with cost} = \text{Cumulative Return - Total commission cost}
\]
Interestingly, the only strategy that outperformed buy-and-hold after accounting for transaction fees was the LSTM.

\section{Discussion}

Both trading approaches are treated as machine learning algorithms and evaluated using a training set and a test set under identical conditions. As expected, the results are positive, meeting the baseline for potential real-world implementation. The technical analysis strategies perform well, yielding cumulative returns comparable to the standard buy-and-hold approach, which is widely favored by fundamental investors due to its simplicity and ease of implementation. These results support the reliability and effectiveness of technical analysis in trading. On the other hand, the LSTM-based strategy, which is grounded in a data-driven and more complex scientific framework, demonstrates superior performance, as anticipated. A natural next step would be to investigate the performance of hybrid models that combine LSTM with traditional strategies—such as EMA or MACD+ADX—to assess whether such integration can enhance predictive power and trading outcomes.

\section*{Author Contributions}
The authors contributed equally to this work. All authors have read and agreed to the published version of the manuscript.

\section*{Funding}
This research received no external funding.

\section*{Data Availability}
The data presented in this study are available at: \url{https://github.com/agarciam/Technical-Analysis-Meets-Machine-Learning-Bitcoin-Evidence/}.


\section*{Conflicts of Interest}
The authors declare no conflicts of interest.

 \bibliographystyle{elsarticle-num} 
 \bibliography{ref}
\end{document}